\documentclass[manuscript]{acmart}

\AtBeginDocument{%
  }

\copyrightyear{2024}
\acmYear{2024}
\setcopyright{rightsretained}
\acmConference[FAccT '24]{The 2024 ACM Conference on Fairness, Accountability, and Transparency}{June 3--6, 2024}{Rio de Janeiro, Brazil}
\acmBooktitle{The 2024 ACM Conference on Fairness, Accountability, and Transparency (FAccT '24), June 3--6, 2024, Rio de Janeiro, Brazil}\acmDOI{10.1145/3630106.3658916}
\acmISBN{979-8-4007-0450-5/24/06}

\begin{document}

\title[Algorithmic Misjudgement in Google Search Results]{Algorithmic Misjudgement in Google Search Results: Evidence from Auditing the US Online Electoral Information Environment}

\author{Brooke Perreault}
\email{bp101@wellesley.edu}
\affiliation{
  \institution{Wellesley College}
  \city{Wellesley}
  \state{Massachusetts}
  \country{USA}
}

\author{Johanna Hoonsun Lee}
\email{hl105@wellesley.edu}
\affiliation{
  \institution{Wellesley College}
  \city{Wellesley}
  \state{Massachusetts}
  \country{USA}
}

\author{Ropafadzo Shava}
\email{rs2@wellesley.edu}
\affiliation{
  \institution{Wellesley College}
  \city{Wellesley}
  \state{Massachusetts}
  \country{USA}
}
\author{Eni Mustafaraj}
\email{emustafa@wellesley.edu}
\affiliation{
  \institution{Wellesley College}
  \city{Wellesley}
  \state{Massachusetts}
  \country{USA}
}

\renewcommand{\shortauthors}{Perreault et al.}

\begin{abstract}
Google Search is an important way that people seek information about politics \cite{SearchAndPolitics2017}, and Google states that it is ``committed to providing timely and authoritative information on Google Search to help voters understand, navigate, and participate in democratic processes.''\footnote{https://elections.google/civics-in-search/} This paper studies the extent to which government-maintained web domains are represented in the online electoral information environment, as captured through 3.45 Google Search result pages collected during the 2022 US midterm elections for 786 locations across the United States. Focusing on state, county, and local government domains that provide locality-specific information, we study not only the extent to which these sources appear in organic search results, but also the extent to which these sources are \textit{correctly targeted} to their respective constituents. We label misalignment between the geographic area that non-federal domains serve and the locations for which they appear in search results as \textit{algorithmic mistargeting}, a subtype of algorithmic misjudgement in which the search algorithm targets locality-specific information to users in different (incorrect) locations. In the context of the 2022 US midterm elections, we find that 71\% of all occurrences of state, county, and local government sources were mistargeted, with some domains appearing disproportionately often among organic results despite providing locality-specific information that may not be relevant to all voters.
However, we also find that mistargeting often occurs in low ranks. We conclude by considering the potential consequences of extensive mistargeting of non-federal government sources and argue that ensuring the correct targeting of these sources to their respective constituents is a critical part of Google's role in facilitating access to authoritative and locally-relevant electoral information.


\end{abstract}

\ccsdesc[500]{Information systems~Web search engines}
\ccsdesc[500]{Information systems~Personalization}

\keywords{algorithm auditing, problematic behavior, google search audit, elections, algorithmic misjudgement}


\maketitle

\section{Introduction}

Electoral denialism---the belief that elections are rigged or that electoral fraud is rampant---in the United States (and other countries, such as Brazil and Peru) is emerging as a significant threat to democracy \cite{cameron2023electoral} and has roots in racial resentment and rising conspiracism \cite{stewart2023public}. Since the early 2000s, internet was regarded as a medium ``uniquely receptive to conspiracism'' \cite{james2000militias}, and in the past two decades, there is little doubt that it has contributed significantly to conspiracism going mainstream \cite{rosenblum2019lot}. Following the 2020 US Presidential Election, ``election rigging'' conspiracies that came to be labeled as ``The Big Lie'' led to the violent attack to the US Capitol on January 6, 2021 \cite{fahey2023big}. While beliefs about electoral fraud are somewhat persistent \cite{enders20212020}, the 2020 electoral denialism movement was different because it was embraced by partisan state legislatures, as reported by the Brennan Center for Justice.\footnote{https://www.brennancenter.org/our-work/research-reports/voting-laws-roundup-june-2023} As a result, a large number of changes to electoral laws (some of them restricting and some of them expanding voting rights), became operational ahead of the 2022 US midterm election.\footnote{https://www.reuters.com/legal/how-new-us-laws-could-trip-up-voters-midterm-elections-2022-11-01/}

In an environment of election denialism on one hand and changing electoral laws on the other, voters must seek accurate, up-to-date, and relevant information about elections and voting. A survey by the Bipartisan Policy Center in Fall 2022 found that voters are ``most likely to look to their state and local election officials, and search engines'' for election information,\footnote{\url{https://bipartisanpolicy.org/blog/new-survey-data-election-information/}} aligning with previous research \cite{SearchAndPolitics2017}. Election officials similarly position themselves as authorities on election information. For example, the National Association of Secretaries of State launched a campaign ahead of the 2020 presidential elections called \#TrustedInfo2020 to ``promote election officials as the trusted sources of election information.'' Operating as \#TrustedInfo during non-federal election years and now \#TrustedInfo2024 ahead of the 2024 presidential elections, NASS states, ``By driving voters directly to election officials' websites, social media pages, and materials, they will be able to receive credible, timely information on each step of the elections process.''\footnote{\url{https://www.nass.org/initiatives/trustedinfo}}

Search engines also position themselves as trustworthy intermediaries for seeking election information. Google states that it is ``committed to providing timely and authoritative information on Google Search to help voters understand, navigate, and participate in democratic processes.''\footnote{https://elections.google/civics-in-search/} In an article about their work to ``support'' the 2022 US midterm elections, Google's Vice President for Trust and Safety writes, ``Our work is centered around connecting voters to the latest election information ...''\footnote{https://blog.google/outreach-initiatives/civics/our-ongoing-work-to-support-the-2022-us-midterm-elections/} and highlights ``features that show data from nonpartisan organizations'' that work to ``connect voters with accurate information about voter registration and how to vote,'' referring to Google Search election information panels that contain aggregated information from \emph{Democracy Works}\footnote{Democracy Works describes itself as ``a nonpartisan, nonprofit organization that collaborates with election officials, leading tech platforms, and world-class partners to drive voter access and participation''. Source:  https://www.democracy.works/about.} about where and how to vote. 

\textbf{Study Motivation: }Given this background, three assumptions motivate the current study: (1) voting-eligible US citizens who search for electoral information using Google's search engine have the right to receive accurate and relevant search results, regardless of where they live; (2) similar to Ballatore et al. \cite{BallatoreLocalness2017}, we acknowledge a spatial dimension to searching and search results, in which users searching for information related to elections and voting deserve \textit{results relevant to their geographic area}, and 3) given the size of the United States and the varying voting laws per state (most of them updated ahead of the 2022 elections), receiving locality-specific information from authoritative  sources is a reasonable expectation. Following these assumptions, we proceed to audit Google Search to determine the extent to which government-maintained web domains, as authoritative sources on election information, are represented in the online environment of electoral information, during the 2022 US midterm elections. Specifically, we analyze where, how often, and how accurately state, county, and local government domains appear in Google's organic results for election-related queries collected in 786 locations across the United States between October and November 2022, in a dataset of 3.45 million search engine result pages (SERPs). Drawing parallels to studies of geotargeting errors in advertisements \cite{GeotargetingError2021}, we label the misalignment between the geographic area that a non-federal government web domain serves (such as a state or county) and the location for which it appears in search results as an instance of \textit{mistargeting}, in which the search algorithm targets locality-specific information to a user in a different (incorrect) location. In this paper, we answer two related research questions:
\begin{itemize}
    \item \textbf{RQ1: }To what extent are government web domains represented in the online environment of electoral information about the 2022 US Midterm Elections, as mediated by Google Search? 
    \item \textbf{RQ2: }To what extent are state, county, and local government web domains correctly targeted to their respective constituents? 
\end{itemize}

\textbf{Findings and Contributions:} 
We find that 40.6\% (1,848/4,556) of domains that appear among organic results are government sources, and that these government sources constitute nearly 40\% of all organic results occurrences. Despite this proportionality, we find that 71\% of the occurrences of state, county, and local government sources were mistargeted, with a handful of locality-specific government domains appearing disproportionately often among organic results, despite providing locality-specific information that may not be relevant to all voters. Despite extensive mistargeting, we find that correctly targeted government sources usually appear in higher ranks, potentially mitigating possible harm inflicted by mistargeting. We conclude by considering how mistargeting factors into Google's role in facilitating access to accurate, authoritative, and locally-relevant electoral information to voters.\footnote{Audit and analysis materials, including query phrases, locations, and scripts, are available on GitHub: \url{https://github.com/credlab/facct24}}

\section{Literature Review}
\subsection{Algorithm Auditing and Problematic Behavior}
Online platforms and other sociotechnical and algorithmic systems are neither neutral nor unbiased entities. As Gillespie \cite{PoliticsofPlatforms2010} argues, the presentation of platforms as open or neutral parties obscures the consequences of deliberate choices platforms make as to what content is published and how content is shown. Through a systematic literature review, Bandy organizes potential problematic behavior of algorithmic systems into four main categories (a) discrimination based on identity or socioeconomic status; (b) distortion or obscuration of reality, (c) exploitation of sensitive user information, and (d) misjudgement and inaccurate classifications \cite{BandyReview2021}. Potential consequences of algorithmic systems become more wide-reaching when considering the interplay between the technical components of algorithmic systems and the social systems in which they exist; one such theme of sociotechnical harm is social systems harms, including how algorithmic systems contribute to the spread of misinformation, disinformation, and malinformation (informational harms) and how algorithmic systems impact governance, democracy, and civil liberties (political and civic harms) \cite{SociotechnicalHarms2023}.

Sandvig et al. \cite{Sandvig2014AuditingA} propose algorithm auditing as a way of studying problematic behavior in Internet platforms and algorithmic systems, particularly problematic behavior that is subtle or unobservable via one-off instances. To date, search engines are among the most audited algorithmic systems, often for \textbf{discrimination} or \textbf{distortion} problematic behavior. Prior research has focused on personalization of search results, including personalization based on logged-in status \cite{HannakPersonalization2013}, browsing history \cite{Le2019GoogleNews, HaimNewsBubble2018}, and location \cite{LocationLocationLocation2015}; composition of search results, including the presence and content of snippets \cite{HuPartisonSnippets2019}, Top stories \cite{LurieTopStories, SearchAsNewsCurator2019}, and other non-organic panels \cite{PersonalizationCompositionRobertson2018}; and prominence of different types of sources among organic results \cite{FischerLocalNews2020, challengeBubble2018, MetaxaSearchMedia}. The pervasive concern in such studies is that the algorithmic choices are leading to a distorted view of reality. Meanwhile, studied to a lesser extent is problematic behavior related to \textbf{algorithmic misjudgement}. As Bandy defines it \cite{BandyReview2021}, misjudgement occurs when an algorithm makes incorrect predictions or classifications, and misjudgement often leads to other types of problematic behavior. Algorithmic misjudgement has been studied in the criminal justice context (eg. recidivism prediction \cite{DuweRecidivism2017}), as well as the advertising context, where algorithms incorrectly infer information about users \cite{BandyReview2021}. One implication of advertising algorithm misjudgement is that purchased advertisements do not reach intended audiences. For example, Bandy and Hecht \cite{GeotargetingError2021} find that geotargeting errors are common in Google Display Network, with users often seeing advertisements not intended for their ZIP code. 
However, they find that severity of such geotargeting errors were minor in the context of advertising goals, with most users living in the intended county, designated market area, or state \cite{GeotargetingError2021}. With this study, we contribute to the literature on algorithmic misjudgement in the context of location-sensitive electoral information.

\subsection{Politics and Search Engines}
Much of previous work on auditing search engines is rooted in a political context, motivated by the potential informational, political, and civic harms that could result from biased search. Epstein and Robertson \cite{EpsteinSEME2015} hypothesized, for example, that biased search results can impact electoral outcomes by shifting voter preferences based on manipulated rankings, as studied in controlled, lab-based experiments \cite{EpsteinSEME2015, EpsteinSupressingSEME2017}. However, algorithm audits in the wild find little evidence for this concern \cite{PAB2018, PersonalizationCompositionRobertson2018}. Findings that support the presence of filter bubbles in Google Search and Google News are also limited \cite{PAB2018, beyondTheBubbleGerman2019, challengeBubble2018, HaimNewsBubble2018, PartisanSearchBehavior2022}. Trielli and Diakopoulos \cite{PartisanSearchBehavior2022} in particular find that while members of different ideological groups use different queries to search for political information, such partisan queries do not lead to a filter bubble effect; rather, they observe a ``mainstreaming effect'', where highly similar results appear even for diverging partisan queries. Moreover, Robertson et al. \cite{RobertsonUsers2023} find that user choice plays a larger role in engagement with partisan or unreliable news than algorithmic curation, further evidence against the idea of partisan filter bubbles.

Given the focus on potential bias or filter bubbles, the query phrases used in most political information audits are intended to capture such phenomena with respect to either the candidates or issues central to an election. Examples include queries related to Donald Trump's inauguration \cite{PersonalizationCompositionRobertson2018}, candidate names \cite{MetaxaSearchMedia}, partisan terms about immigration \cite{Le2019GoogleNews}, socio-political themes, \cite{challengeBubble2018}, and crowd-sourced queries from users about senate candidates \cite{PartisanSearchBehavior2022}. Most of the audits that focus on candidate names are conducted during election periods, including the 2016 US Congressional Elections \cite{Metaxas2017ManipulationOS}, the 2018 US Midterms \cite{PartisanSearchBehavior2022, MetaxaSearchMedia}, and 2020 US Presidential Primaries \cite{UrmanChange2022, kawakami2020media}.  Diakopoulos et al. \cite{diakopoulos2018vote} and Lurie and Mulligan \cite{LurieBreakdown2021} similarly study various information panels that appear in Google search results when searching for candidate or representative names, with the latter finding that featured snippets are often likely to mislead searchers by presenting incorrect or incomplete information about a representative. Lurie and Mulligan contextualize this behavior as ``algorithmic breakdown'', and identify place name ambiguity as an important contributor to such breakdown, wherein the algorithmic system is unable to discern a particular location (and the respective locality-relevant information, like the representative of a congressional district) from an ambiguous place name, such as a county and city that share the same name. Diakopoulos et al. \cite{diakopoulos2018vote} additionally study the composition of organic results for queries of Democrat and Republican candidates ahead of the 2016 US presidential election, finding that official sources, such as campaign websites, were clustered at the top of organic results in high ranking positions. Most of these studies focus on query phrases selected by researchers in various ways. Mustafaraj et al. \cite{VoterCenterAudits2020} discuss that voters have information needs not previously captured in such audits. Building upon prior literature, we create an election-relevant query phrases dataset composed of popular queries from Google Trends and queries formulated by users as reported in \cite{VoterCenterAudits2020}.

\subsection{Source Geoprovenance}
As Ballatore et al. \cite{BallatoreLocalness2017} describe, search engines ``increasingly mediate not just information but also spatial knowledges and experiences,'' as users rely on search engines to inform them about destinations, where to shop, or other ways to interact with their ``lived everyday geographies'' - including participating in civic processes, such as voting.  Moreover, \textit{where} information originates from, particularly if information originates from local or non-local sources, can impact how people learn about and interact with places near them, to the extent that local community members and non-community members view their locality differently \cite{SenGeoprovenance2015}. The geographic origin of information, or geoprovenance \cite{SenGeoprovenance2015}, has largely been studied in places where volunteered geographic information (e.g. geotagging) is accessible, such as Wikipedia \cite{SenGeoprovenance2015} and Twitter \cite{JohnsonGeotaggedLocalness2016}. Ballatore et al. \cite{BallatoreLocalness2017} study the extent to which different country variants of Google Search direct users to locally produced information about places, the first such audit study of search engines. Using a geoprovenance inference algorithm developed by Sen et al. \cite{SenGeoprovenance2015}, they identify the geoprovenance of a URL at the country level and compare URL geoprovenance to the country of search, finding a greater degree of local content for wealthy and well-connected countries. This study uses the concept of geoprovenance of a URL to study algorithmic misjudgement in search results.  We audit not only the extent to which government sources are present in search results, but also the extent to which state, county, and local government sources are shown to the geographic areas which they serve. 

\section{Data Collection}
An audit comprises a set of inputs and the corresponding outputs. Our inputs are the query phrases and the geographical locations where Google searches were performed. The outputs are the search engine results pages (SERPs) generated for each query-location pair. We describe how we selected the queries and locations and how we performed the audit.

\subsection{Selection of Queries}
In line with previous research about centering voters in algorithm audit methodology \cite{VoterCenterAudits2020}, we developed a list of queries aimed at capturing the diverse information needs of voters who use Google to search for election information. 

In early October 2022, we collected Top and Rising related terms from Google Trends for the seed queries ``midterm elections", ``voting near me'', and ``ballot'', as well as from the Google Trends category ``United States election, 2022 - General election''. The location of Google Trends was the United States. From the received results, we excluded terms that were irrelevant to the topic of elections; terms that included the names of political figures or candidates; and any terms that included specific states, state abbreviations, or localities so that queries generalize across locations. 


A review of the query list after these steps revealed a lack of general queries about candidates and key voter issues, as well as a lack of biased and/or partisan queries. To fill these gaps, we augmented our query list by manually selecting queries formulated by eligible US voters included in the Voter Searches dataset in \cite{VoterCenterAudits2020}. Lastly, a few additional seed queries were added related to the topic of abortion, which was one of the top voting issues in the 2022 election.\footnote{https://www.pewresearch.org/politics/2022/08/23/abortion-rises-in-importance-as-a-voting-issue-driven-by-democrats/} These processes resulted in a final list of 161 queries, of which 74 were from Google Trends. 

Queries were reviewed qualitatively by one author to develop three major themes of a) \textit{Voting Logistics} (in which queries are primarily about logistical information related to voting); b) \textit{Candidates and Issues} (in which queries are primarily about candidates and candidates' opinions, stances, and/or leanings)\footnote{The category \textit{Candidates and Issues} includes 9 queries about abortion, a key voting issue in the 2022 US midterms. While these 9 queries could constitute their own category, we found that separating ``Candidates" and ``Issues" queries did not lead to meaningful changes in any results with respect to mistargeting rates, as government sources appear very minimally for these 9 queries. Thus, it was sufficient to lump these queries in with queries about candidates and candidate stances. Future work may study mistargeting for a more expansive set of queries representing different key voting issues.} and c) \textit{General Election Topics}, (where queries are primarily general inquiries about the midterm election, such as what midterms are and when they occur). Three labelers then independently labeled all queries into these three categories. The majority label was used as the final label per query. Table \ref{tab:queries} shows the query distribution by category and some examples for each category. 

\begin{table}[h!]
\centering
\begin{tabular}{lll}
\hline
\textbf{Query Category}          & \textbf{Example Queries}                                                 & \textbf{n} \\ \hline
Voting Logistics        & voting early near me, sample ballot for my county, what district am i in & 68         \\
Candidates and Issues   & candidate midterm info, candidates for my district, candidate opinions& 50         \\
General Election Topics & 2022 midterm, mid term elections, should i vote,                         & 43         \\ \hline
\end{tabular}
\caption{The audit used 161 queries that cover three major themes related to logistics of voting, candidates, and general inquiries about midterms. The query categories were assigned by three independent labelers. }
\label{tab:queries}
\end{table}

\subsection{Selection of Locations}
We selected locations for data collection to correspond to U.S states and congressional districts whose House or Senate midterm races were classified as toss-ups, leaning Democrat, and leaning Republican, according to the Cook Political Report.\footnote{ \url{https://www.cookpolitical.com/ratings/senate-race-ratings} and \url{https://www.cookpolitical.com/ratings/house-race-ratings}} We identified 30 congressional districts with competitive House races across 23 states, as well as 10 states with competitive Senate races; 3 states had only competitive Senate races, making for a total of 26 unique states.

For house races, because congressional district borders do not align with county borders, we first manually compiled a list of counties within the relevant congressional district, according to Wikipedia,\footnote{Eg. \nolinkurl{https://en.wikipedia.org/wiki/California's_27th_congressional_district}} including only the counties that were described as having most or all of their localities within the congressional district. Next, we identified the associated Federal Information Processing System (FIPS) Codes for each county\footnote{https://transition.fcc.gov/oet/info/maps/census/fips/fips.txt} and used these codes to filter 2020 U.S. Census data\footnote{Dataset titled ``Incorporated Places and Minor Civil Divisions Datasets: Subcounty Resident Population Estimates: April 1, 2020 to July 1, 2021 (SUB-EST2021)'' from https://www.census.gov/data/tables/time-series/demo/popest/2020s-total-cities-and-towns.html} to include localities from the relevant counties only. Additionally, we filtered locations to only include cities, towns, and townships, as identified on the Census. 

To sample locations, we identified the median population of possible locations, and selected (up to) 15 locations below the median population and (up to) 15 locations above the median population, as well as the location with the highest population. Some states (for example, Nevada, which had only a competitive Senate race) had fewer than 31 cities, towns, and townships from the relevant counties. This process resulted in 872 locations across 26 states. Google Places API was used to get latitude and longitude coordinates for each of the locations. This information is needed to change the location of the Chrome browser to the desired location for the audit,
as shown next.

\subsection{Scraping Audit and SERP Validation}
\begin{figure}[h!]
    \centering
    \includegraphics[width=0.5\linewidth]{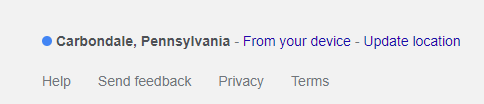}
    \caption{The script automatically updates the location with latitude and longitude coordinates so that the search location is ``from your device'' rather than IP address. Previous audits \cite{LocationLocationLocation2015} have tested and confirmed that search results are personalized based on provided coordinates, rather than the IP address.}
    \label{fig:location_box}
\end{figure}

We use the Python-scripted Selenium web driver to automate the process of querying Google Search, changing the geolocation, and saving the SERPs. For each query, our custom script opens a blank-slate of the Chrome browser, enters the query, and scrolls to the bottom of the page to update the location using latitude and longitude coordinates (Figure \ref{fig:location_box}). The page reloads, displaying the results for the new location, and the entire HTML page is stored. 

\vspace{-10pt}

SERPs were collected for 30 days, starting on October 18, 2022 and ending November 18, 2022. Data were not collected on November 14 and 15. Data were collected using 12 desktop computers, with 2 to 3 states collecting per computer. Because of the large amount of locations, data collection started each day at 2am EST and continued until the evening. The order of locations was not changed daily, so SERPs per query were collected at approximately the same time within each location, and locations were completed sequentially. Due to random hardware or software failure, some SERPs failed to collect. 3,838,128 SERPs were collected in total (92.3\% collection rate). 

To validate that locations were changed correctly, we extracted the text of the location box on each SERP and compared it to the intended location. The validation process revealed two types of errors: (1) the location of a SERP failed to update properly for unknown reasons, and (2) the latitude and longitude coordinates from the Google Places API did not exactly match the actual intended location. To resolve these issues, we remove all 86 locations that had errors and conduct analysis on 3,455,822 SERPs from the remaining 786 locations across 25 states. 

\section{Methods}
Since our research questions revolve around the presence of government-maintained websites in SERPs, we describe in this section how we extracted their occurrences, labeled them according to the government type and geographic area, and defined when they are mistargeted with respect to the search location.

\subsection{Extracting domains}
SERPs are composed of organic results (the blue links) and non-organic results that include top stories, videos, knowledge panels, and, during election season, election information panels. This study focuses on organic results only, and in particular the domains that occur in organic results.
22,750 unique links were parsed from 31,310,263 organic results. Once all such links were extracted from all SERPs, they were cleaned to their root domain; however, given our focus on various government sources, in some cases we maintain the subdomain and avoid aggregating to the root, particularly if the subdomain is a county or town website. For example, while \texttt{clintonco.illinois.gov} is a subdomain of the state government website \texttt{illinois.gov}, we maintain the county government website. 4,556 unique domains were identified. 
For each SERP, we additionally record the rank of the organic results. In determining rank, we do not account for the presence of non-organic panels that may be on the page; rank is based only on the organic results.


\subsection{Identifying Government Websites}
To identify government websites, we make use of two repositories maintained by the US government: (1) a list of nearly 10,000 .gov domains maintained by the Cybersecurity and Infrastructure Security Agency (CISA),\footnote{https://github.com/cisagov/dotgov-data/blob/main/current-full.csv} and (2) a list of  9,200 government-maintained websites outside of the .gov and .mil domains, maintained by the US General Services Administration (GSA).\footnote{https://github.com/GSA/govt-urls/blob/main/1\_govt\_urls\_full.csv} Both datasets include information about the type of government domain, such as state, county, or local, the agency that maintains the domain, and some geographic information about the organization that the domain serves (such as the state or county). 726 domains were found in the CISA repository and 716 domains from the GSA repository.
Because of how links were cleaned and aggregated, some .gov websites, particularly those corresponding to towns, cities, or counties, were not included in these repositories because they were subdomains of state root domains. We manually examined remaining websites ending with .gov and included any that correspond to a town, city, or county government, manually filling in relevant information about government type and geographic area. 

Additionally, we identified some county and city websites using the .us domain that were not included in the GSA repository; because the .us domain is not strictly governmental, we manually reviewed all websites with the .us domain and included only those which corresponded to a county or city. 362 domains were added by manual review. In total, 1,848 websites were identified, which we call ``government domains'' or ``government sources'.' 
The distribution of domains across government types is shown in Table \ref{tab:govt_domains}. 

\begin{table}[t!]
\centering
\begin{tabular}{llr}
\hline
\textbf{Government Domain Type} & \textbf{Examples (geoprovenance)}             & \textbf{n} \\ \hline
Federal                         & whitehouse.gov (United States)                & 44         \\
State                           & az.gov (Arizona)                              & 160        \\
County                          & cookcountyclerkil.gov (Cook County, Illinois) & 697        \\
Local                           & cityofmaywood.com (Maywood, California)       & 806        \\
Representative                  & bishop.house.gov (Georgia, 2nd District)      & 137\\
Senator                         & portman.senate.gov (Ohio)                     &            3\\
Native                          & nativefederation.org (Alaska)                 & 1          \\ \hline
\end{tabular}
\caption{1,848 government sources were identified from the data set of 4,556 domains (40.6\%). Government sources were labeled for type and geoprovenance. We focus our analysis on state, county, and local government sources, as senator, representative, and Native sources show up minimally, composing less than 0.1\% of all organic results.}
\label{tab:govt_domains}
\end{table}

\subsection{Measuring Mistargeting}
We label misalignment between the geographic area that non-federal domains serve and the locations for which they appear in search results as \textit{mistargeting}, a type of algorithmic misjudgement in which the search algorithm targets locality-specific information to users in different (incorrect) locations. Mistargeting can be studied at varying degrees of granularity.

\subsubsection{Domain level} First, we examine mistargeting with respect to government domains and where they appear. To do so, we first identify how many government domains that appeared in organic results are from states in which we did not collect data; these are domains which are always mistargeted. Next, for domains originating in states for which we collected data, we calculate the proportion of times the domain appeared in the appropriate state and the proportion of times the domain was mistargeted (at the state-level). This gives insight into whether there are some locality-specific government domains that Google shows often across locations and are frequently mistargeted.

\subsubsection{Aggregate Organic Results}
We also study mistargeting by looking at all organic results in aggregate and and by parsing differences in mistargeting by query categories, domain type, and state. Every organic result is labeled as to whether it is a government domain, and if so, we compare the domain's geoprovenance (the geographic area it serves, such as a state, county, or city) to the search location. This gives us a more precise understanding of mistargeting than at the domain level, in which we only compare states. Organic results are labeled as either correct or mistargeted. Mistargeted results for county and city domains are also labeled as either in-state, if they serve a different location in the same state as the search location, or out-of-state, if they do not, which functions as an indicator for the severity of mistargeting. For each query category, and across all queries, we calculate the proportion of organic results composed of \textit{all} non-federal government sources and the proportion of results composed of\textit{ correctly targeted }government results. Mistargeting rate is calculated by subtracting the proportion of correctly targeted from the total proportion, and dividing by the total proportion. We similarly calculate the proportion of results that are mistargeted but in-state, which gives insight into the nature of mistargeted results.

\subsubsection{SERP Level}
Lastly, we quantify mistargeting at the SERP level by calculating descriptive statistics about the number of government sources that appear per SERP, as well as descriptive statistics about the rank of correctly and incorrectly targeted government results. Since users heavily rely on ranking of search results \cite{Pan2007}, this informs the severity of misjudgement with respect to a user.

\section{Results}

\subsection{RQ1: To what extent are government sources represented in the online environment of electoral information? }
 Of the 4,556 domains that appeared in organic results, 1,848 were identified as a government domain (40.6\%). 
Overall, links belonging to these 1,848 government domains compose 39.97\% of all organic results captured in the 3.45 million SERPs (31.3 million organic results) analyzed. Non-federal government domains compose 32.89\% of all organic results. However, the extent to which government domains appear depends on the type of query, as well as the type of government source as shown in Table \ref{tab:govt_appearances}. In particular, government sources constitute 60.55\% of organic results for \textit{Voting Logistics} queries, with state government sources composing half of such appearances. In contrast, government sources are underrepresented among organic results for \textit{General Election Topics} and \textit{Candidates and Issues} queries, composing  just 26.36\% and 23.15\% of organic results, respectively.

\begin{table}[t!]
\centering
\begin{tabular}{lrrrr}
\hline
                            & \textbf{Voting Logistics} & \textbf{General Election Topics} & \textbf{Candidates and Issues} & \textbf{All Queries} \\ \hline
\textbf{\% Federal}         & 4.85                      & 9.1                              & \multicolumn{1}{r|}{8.44}      & 7.08                 \\
\textbf{\% State}           & 30.06                     & 11.99                            & \multicolumn{1}{r|}{11.87}     & 19.67                \\
\textbf{\% County}          & 22.91                     & 4.75                             & \multicolumn{1}{r|}{2.66}      & 11.86                \\
\textbf{\% Local}           & 2.62                      & 0.52                             & \multicolumn{1}{r|}{0.18}      & 1.31                 \\ \hline
\textbf{\% Total excl. Federal} & 55.7                      & 17.26                            & \multicolumn{1}{r|}{14.71}     & 32.89                \\
\textbf{\% Total}           & \textbf{60.55}                    & 26.36                            & \multicolumn{1}{r|}{23.15}     & \textbf{39.97}                \\ \hline
\end{tabular}
\caption{Government domains compose 39.97\% of organic results overall, with state and county government domains constituting the majority of government domain appearances, while local government domains are minimally represented in organic results.
The degree to which government domains appear depends largely on the query type: government sources compose 60\% of organic results for \textit{Voting Logistics} queries but only around a quarter for \textit{General Election Topics} and \textit{Candidates and Issues}. }
\label{tab:govt_appearances}
\end{table}

 Across all queries, local government sources are minimally represented among organic results. The prominence of state-level government sources compared to other government sources may be explained by the key roles that state-level actors play in elections: while the composition of election administrations vary by state, elections in each state are overseen by a chief election officer (such as a secretary of state, a governor, or a board), and their responsibilities include enforcing election laws, administering a statewide voter registration database, and other duties related to election logistics.\footnote{https://www.ncsl.org/elections-and-campaigns/election-administration-at-state-and-local-levels} However, their dominance is striking. While state government websites make up 3.5\% of all websites (160 out of 4,556), they compose 30.06\% of \textit{Voting Logistics} results and 19.67\% of overall results.

On a SERP level, \textit{Voting Logistics} queries had a median of 6 government results, with medians of 3 state-level and 2 county-level government sources, aligning with the aggregate proportions in Table\ref{tab:govt_appearances}. In contrast, \textit{General Election Topics} queries had a median of 2 government source per page and \textit{Candidates and Issues} queries had a median of 1. 

\subsection{RQ2: To what extent are state, county, and local government web domains correctly targeted to their respective constituents?}

\subsubsection{Domains}
At the domain level, 174 of the 1,804 (9.6\%) non-federal government sources represent a state for which we did not collect data; in other words, these sources were always mistargeted. This includes 71 local-level, 53 county-level and 48 state-level government domains, along with 2 senator websites. Appearances from these domains constitute 13.65\% of appearances of non-federal government sources, or about  4.4\% of all organic results. The remaining 1,630 non-federal government sources match, at the state level, the states for which we collected SERPs. 1,268 of these domains always show up in the appropriate state (although not necessarily in the appropriate county or locality, for county and local sources). While this is the majority of non-federal government domains (70.3\%), only 4.96\% of all organic appearances by non-federal government sources can be attributed to these domains. 

\begin{table}[t!]
\centering
\begin{tabular}{lllrrc}
\hline
\textbf{Domain} &  \textbf{Type}&\textbf{\begin{tabular}[c]{@{}l@{}}State\\ Domain Serves\end{tabular}} & \textbf{\begin{tabular}[c]{@{}c@{}}\% Appearances\\ Correct State\end{tabular}} & \textbf{\begin{tabular}[c]{@{}c@{}}\% Appearances\\ Other States\end{tabular}} & \textbf{\begin{tabular}[c]{@{}c@{}}Collected data\\ in state?\end{tabular}} \\ \hline
pa.gov          &  State&Pennsylvania                                                           & 55.25                                                                           & 44.75                                                                          & Yes\\
ca.gov          &  State&California                                                             & 61.98                                                                           & 38.02                                                                          & Yes\\
sos.state.tx.us &  State&Texas                                                                  & 18.86                                                                           & 81.14                                                                          & Yes\\
wi.gov          &  State&Wisconsin                                                              & 26.30                                                                           & 73.69                                                                          & Yes\\
wakegov.com     &  County&North Carolina                                                         & 5.78                                                                            & 94.52                                                                          & Yes\\
lavote.gov      &  County&California                                                             & 25.79                                                                           & 74.21                                                                          & Yes\\
ncsbe.gov       &  State&North Carolina                                                         & 34.45                                                                           & 65.55                                                                          & Yes\\
georgia.gov     &  State&Georgia                                                                & 28.28                                                                           & 71.71                                                                          & Yes\\
maryland.gov    &  State&Maryland                                                               & 0                                                                               & 100.00                                                                              & No\\
votebrevard.gov &  County&Florida                                                                & 5.83                                                                            & 94.17                                                                          & Yes\\ \hline
\end{tabular}
\caption{These 10 domains constitute 35.96\% of all non-federal government organic results, which is 11.8\% of all organic results. While collecting from more locations in certain states (such as Pennsylvania and California) increases the number of times these domains occur, all of these domains appear often in states for which they do not serve, indicating frequent mistargeting. \texttt{maryland.gov} is one of 174 non-federal sources that appear despite no data being collected for its respective state. }
\label{tab:domain_apps}
\end{table}

Table \ref{tab:domain_apps} shows the 10 non-federal government domains which make up the highest proportion of organic results, and together compose 35.96\% of appearances for non-federal domains, or 11.8\% of organic results overall. Because the number of locations for which data was collected varied per state, it is not appropriate to directly compare the frequency with which these domains occur to each other. However, it is clear that these domains suffer from high rates of mistargeting, with many appearing more often in other states than the states for which they serve. Notably, \texttt{wakegov.com}, a county government site for Wake County, North Carolina, and \texttt{votebrevard.gov}, the Supervisor of Elections site for Brevard County, Florida, are mistargeted over 94\% of the time. Overall, 84 domains were mistargeted more often than they were correctly targeted, and they compose 44\% of non-federal organic results (14.45\% of all organic results). In contrast, 245 domains are correctly targeted more often than they are mistargeted, and they compose 41\% of non-federal organic results (13.5\% of all organic results). Thus, while the majority of non-federal government sources are always correctly targeted at the state-level, these sources make up only a minority of all organic results; a relatively small subset of domains are mistargeted more often than they are correctly targeted, and these mistargeted results show \textit{frequently} among organic results.
 \begin{table}[t!]
\centering
\begin{tabular}{lrrrr}
\hline
                           & \textbf{Voting Logistics} & \textbf{General Election Topics} & \textbf{Candidates and Issues} & \textbf{All Queries} \\ \hline
\textbf{\% State Correct}  & 14.5                      & 4.1                              & \multicolumn{1}{r|}{3.0}         & 8.2                  \\
\textbf{\% County Correct} & 1.89                      & 0.42                             & \multicolumn{1}{r|}{0.11}      & 0.96                 \\
\textbf{\% Local Correct}  & 0.63                      & 0.13                             & \multicolumn{1}{r|}{0.03}      & 0.31                 \\ \hline
\textbf{\% Total Correct}  & 17.02                     & 4.65                             & \multicolumn{1}{r|}{3.14}      & 9.47                 \\ \hline
\end{tabular}
\caption{Percent of \textit{all organic results} that are correctly targeted to the appropriate locality for state, county, and local sources; the denominator here is the number of all organic results (about 31 million). While 32.89\% of all organic results can be attributed to locality-specific government sources, only 9.47\% of organic results are composed of correctly targeted government sources.}
\label{tab:correct_appearances}
\end{table}
\subsubsection{Mistargeting by Query Category and State}
Table \ref{tab:correct_appearances} shows the percent of all organic results that are correctly targeted to the appropriate locality, and Table \ref{tab:mistargeting} shows the mistargeting rates for each state, county, and local government sources for different query categories. Across all SERPs, 71\% of all appearances of locality-specific government websites in organic results were mistargeted. Mistargeting is highest among \textit{Candidates and Issues} queries at 78.61\%. \textit{Voting Logistics} queries, which see the highest proportion of government results by far, have the lowest mistargeting rate of the 3 query categories, albeit still 69.4\%. 
\begin{table}[]
\centering
\begin{tabular}{lrrrr}
\hline
                               & \textbf{Voting Logistics} & \textbf{General Election Topics} & \textbf{Candidates and Issues} & \textbf{All Queries} \\ \hline
\textbf{\% State Mistargeted}  & 51.75                     & 65.8                             & \multicolumn{1}{r|}{74.73}     & 58.29                \\
\textbf{\% County Mistargeted} & 91.73                     & 91.12                            & \multicolumn{1}{r|}{95.7}      & 91.94                \\
\textbf{\% Local Mistargeted}  & 75.92                     & 75.06                            & \multicolumn{1}{r|}{81.79}     & 76.08                \\ \hline
\textbf{\% Total Mistargeted}& 69.4                      & 73.05                            & \multicolumn{1}{r|}{78.61}     & 71.18                \\ \hline
\end{tabular}
\caption{Percent of \textit{government results} that were mistargeted. Mistargeting rate is calculated by subtracting the proportion of correctly targeted (Table \ref{tab:correct_appearances}) from the total proportion (Table \ref{tab:govt_appearances}), and dividing by the total proportion. . Overall, 71\% of locality-specific government organic results were mistargeted, with the mistargeting rate being highest for county government sources at nearly 92\%.}
\label{tab:mistargeting}
\end{table}
Across all queries, county-level government sources suffer from the highest levels of mistargeting, at 91.9\%. State-level government domains have the lowest mistargeting rate of the locality-specific sources at about 58.3\%. Mistargeting of state-level government sources is lowest for \textit{Voting Logistics} queries at 51.75\%, notable since state-level sources are the most commonly appearing government source in this query category.  Lastly, local government sources suffer from a mistargeting rate of 76\%, unfortunate considering the minimal appearances of local government sources in organic results. 
\begin{figure}[b!]
    \centering
    \includegraphics[width=0.5\linewidth]{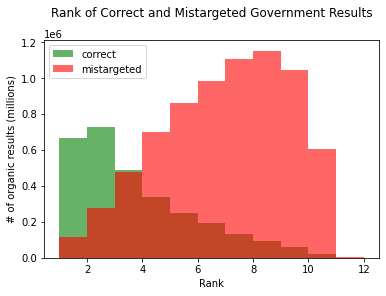}
    \caption{While only 28.8\% of non-federal government domains are correctly targeted, these results usually appear in higher ranks among organic results. On the 43.7\% of SERPs that have both correctly-targeted and mistargeted results, the correctly targeted result appeared above the mistargeted result 83.9\% of the time. Thus, while mistargeting rates are high, their severity, within the context of impact on a user, may be low, given that users often rely on top ranking results. }
    \label{fig:rank_histogram}
\end{figure}

\begin{figure}[hb!]
     \centering
     \includegraphics[width=0.6\linewidth]{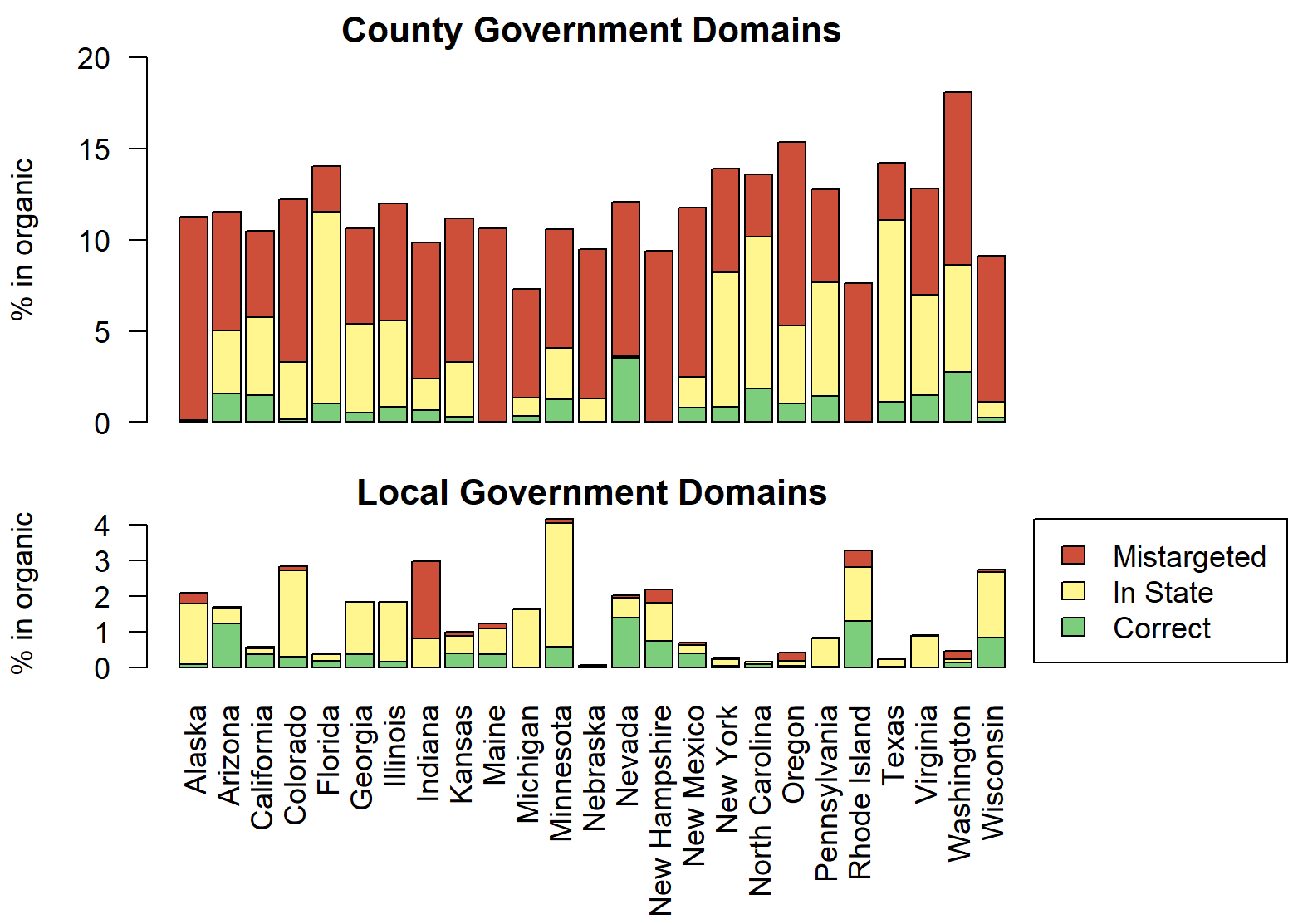}
     \caption{Much of the mistargeting of local government sources (and county government sources, to a lesser extent) can be explained by the appearance of sources that serve other localities in the same state as the search location. This indicates a \textit{lack of precision} with targeting locality-specific government domains to their specific audience when relying only on geocoordinates to represent user location.}
     \label{fig:misjudgement}
 \end{figure} 

\subsubsection{Mistargeting Severity}
While mistargeting rates are high, particularly for county and local government domains, it is useful to distinguish between error rate and error severity. Error severity can be assessed in two ways: 1) the rank of mistargeted organic results, 2)  how close mistargeted results are to the correct location.

With respect to rank, while only 28.8\% of non-federal government sources are correctly targeted, these correctly targeted results generally appear in higher ranks among organic results than mistargeted results (Figure \ref{fig:rank_histogram}). The median rank of a correctly-targeted government result was 3, while the median rank of a mistargeted government result was 7. 
At the SERP level, 5.6\% of SERPs analyzed had only correctly targeted results, 19.9\% of SERPs had only mistargeted results, 43.7\% of SERPs had both correctly-targeted and mistargeted government results, and the remaining 30.7\% of SERPs had no subnational government source. Thus, just under half of SERPs (49.3\%) had at least one correctly targeted result. 
Moreover, for the 43.7\% of SERPs that contained both correctly-targeted and mistargeted government results, the first correctly-targeted result appeared above the first mistargeted result 83.9\% of the time. Lastly, a mistargeted government result appeared in the first organic position for only 3.3\% of SERPs analyzed, while a correctly-targeted government result appeared in the first organic position for 19.3\% of SERPs. 

Given that users tend to focus on search results in the top positions  \cite{Pan2007}, these findings suggest that while mistargeting rates are high, their severity, within the context of impact on a user, may be low. Additionally, these findings may indicate that the search algorithm prioritizes non-federal government sources in ranked results only when there is a high likelihood that the source geoprovenance matches the user's locations; otherwise, non-federal government sources appear in lower ranked positions, and other content (such as a news or non-profit organizations) is given the top ranks.

With respect to the nature of mistargeted results, a high percentage of mistargeting for local government sources can be explained by the appearance of sources that serve other localities in the same state as the search location. While the overall mistargeting rate for local government sources is 76\%, 64\% of local government results are for other localities in the correct state. Figure \ref{fig:misjudgement} visualizes this phenomenon; in most states, nearly all of the mistargeted local government results are from other local government sources in the respective state. 

A similar phenomenon with targeting county-level government sources is observed, though to a lesser degree and with more variation between states. These findings imply a\textit{ lack of precision} with targeting locality-specific government sources to the appropriate audience when using general queries that do not specify location information. While a similar lack of precision has been observed with geotargeting of Ads on Google \cite{GeotargetingError2021}, the consequences of this lack of precision is greater in the context of ``low information voters''\footnote{https://en.wikipedia.org/wiki/Low\_information\_voter} searching for election information as it relates to their locale, given that other county or locality information is likely not be applicable to them.

\section{Discussion, Limitations, and Future Work}

\subsection{Discussion} This paper studies the extent to which local, county, and state government sources appear in election-related search results, as well as the extent to which these sources are correctly targeted to the locations which they serve. We label misalignment between the geographic area that non-federal domains serve and the locations for which they appear in search results as \textit{mistargeting}, a subtype of algorithmic misjudgement in which the search algorithm targets locality-specific information to users in different, incorrect locations.

\subsubsection{Overrepresentation of Some Locality-Specific Government Sources} With respect to our first research question, we find that 40.6\% (1,848/4,556) of websites that appeared in organic results are government sources, and combined, these sources compose 39.97\% of all organic results. This might look like an excellent outcome, given the almost perfect matching between the proportion of domains and the proportion of their occurrences in the search results. However, we find that that some locality-specific government domains are over represented and frequently mistargeted in organic results. About 1/3 of all occurrences of non-federal government sources can be attributed to ten domains, with these domains frequently appearing in locations for which they do not serve (see Table~\ref{tab:domain_apps}). This phenomenon is reminiscent of source concentration and mainstreaming effects found in previous work for news sources \cite{PartisanSearchBehavior2022, SearchAsNewsCurator2019}. While it might not be harmful to an individual to read national news from New York Times more than from Baltimore Sun (although, collectively, source concentration has led to the so-called 'news desert' phenomenon \cite{abernathy2018expanding}), for elections, all out-of-state web domains might be irrelevant to a voter, when it comes to their voting information. 

\subsubsection{Extensive Mistargeting in Low Ranks} Furthermore, we find extensive mistargeting of locality-specific government sources: 71.18\% of all occurrences of non-federal government sources were mistargeted, with county-level government sources having the highest mistargeting rate at 91.94\% (Table \ref{tab:mistargeting}). However, we also find that correctly targeted results generally appear in higher ranks among organic results than mistargeted results (Figure~\ref{fig:rank_histogram}), and for the 43.7\% of SERPs that contained both correctly targeted and mistargeted government results, the first correctly targeted result appeared above the first mistargeted result 83.9\% of the time. In the absence of user behavior data, these findings are encouraging that the severity of harm to a user due to mistargeted results may be low, since low-ranking mistargeted results are usually accompanied by high-ranking correctly targeted results.

\subsubsection{Government Sources, SERP Composition, and Consequences of Mistargeting} The study of mistargeted government results in election SERPs is motivated by how voters and election officials position election officials and government sources as important, trusted conveyors of election information.\footnote{See Introduction; \url{https://bipartisanpolicy.org/blog/new-survey-data-election-information/} and \url{https://www.nass.org/initiatives/trustedinfo}}  It is not our aim to claim that government sources are the most (or only) relevant type of source for electoral information. Indeed, it is conceivable that higher quality, non-government sources containing relevant information may be more beneficial to a user than lower quality, correctly-targeted government sources. Exploring the extent to which government sources are quality sources for electoral information is beyond the scope of this paper. However, regardless of whether government sources should be considered as the best and highest quality, our audit shows not only that locality-specific government sources \textit{are} showing in SERPs for different types of queries (even if minimally), but also that these results are often not showing up \textit{in the correct places}.
Questioning whether a locality-specific government source is correctly targeted to its constituents perhaps precedes questions about whether this source type is most favorable for a user. At best, mistargeting of government sources among organic search results constitutes wasted space on a SERP that could otherwise be used to provide voters with relevant electoral information. At worst, mistargeting constitutes a failure in Google's ability to facilitate (easy) access to accurate, authoritative, and locality-specific electoral information to voters who turn to the search engine with their election-related information needs.


Google states that it is committed to providing voters with authoritative information about US elections. Our focus on mistargeting of government sources among organic results reveals room for improvement in this regard. Showing locality-specific government sources in organic results \textit{and} ensuring that these sources are correctly targeted to their respective constituents are necessary criteria for ensuring that Google Search can appropriately provide users across different locations access to authoritative \textit{and} locally-relevant election information.


\color{black}


\subsection{Limitations}
There are limitations to our study. First, we focus exclusively on government web domains in search results, but in reality there are other types of information that users encounter in search results, including results from other types of sources and curated information panels on the SERP. 
During the 2022 US midterm elections in particular, Google populated election-related SERPs with election information panels, many of which contained curated, state-specific information; an analysis of these panels is necessary to get a complete understanding of how locality-specific electoral information is surfaced among search results, but this is left to future work. Additionally, our audit was conducted when there were around 10 organic results per SERP, though since December 2022 (after our data collection), Google has moved to ``continuous scrolling'', which loads up to 60 results at once.\footnote{https://www.theverge.com/2022/12/6/23495973/google-search-desktop-continuous-scrolling}

Second, we rely on geocoordinates to serve as a proxy for locations of real users and do not study how including location information in the search query (e.g. ``voting near me Boston, MA'') impacts the composition and mistargeting of government search results. We also do not study whether targeting based on geocoordinates differs from IP-based targeting, nor do we consider how user profiles impact targeting. It is conceivable that IP-based targeting and/or user profiles lead to more location precision and thus less mistargeting among organic results. 
Furthermore, while the queries come from Google Trends and real users, we do not study how users actually interact with government sources in search results. 

We analyze data from 786 locations across 25 states, a size which is sufficient for statistical analysis but does not capture all of the United States. Additionally, we do not analyze how mistargeting varies by search location features (including the extent to which the congressional election for a location is competitive) nor by date. Finally, this study only focuses on the United States, given that electoral laws and election authorities are different state by state, making this kind of analysis meaningful. In other countries, where there is a single election authority, the needs and assumptions might be different. Indeed, it is unclear if these results can be generalized to other countries with similar political systems to the United States. In addition to the similarity of political systems, one would have to consider the strength of a country’s online government and electoral information infrastructure when generalizing results. 

\subsection{Future work}

One challenge that we encountered when analyzing our results was finding accurate information about government websites (as discussed in Methods). However, government websites were only 40.6\% of the entire dataset of websites. The rest was composed of news organizations, non-profit civic organizations, companies, universities, individual blogs, and more. There are no high-quality and freely available databases that provide detailed information on web domains (their nature and what audiences they serve). Nuanced audits of search engine results need such information; therefore, methods to create and maintain such databases need to be developed. (We tried to use some news related curated lists for other studies, such as \cite{FischerLocalNews2020}, but their overlap with our dataset was not high.)

A scraping audit is useful for controlling the inputs to the algorithmic system, so that we can systematically compare results, but it does not involve real voters with real information needs, that is, it lacks ecological validity. Currently, there are efforts underway like the National Internet Observatory\footnote{https://nationalinternetobservatory.org/} that will enable researchers to carry out crowdsourced audits, similar to what was used in \cite{RobertsonUsers2023}. Future audits can then address both issues that are central to our audit: what queries do users formulate (and reformulate) to find information about elections and with what results do they engage? For example, do they click on Google's election boxes, do they click on their own local/county/state websites, or do they visit non-governmental sources? Additionally, such audits might allow us to find out if Google Search is bypassed entirely and users go directly to their trusted sources. 


\subsection{Ethical Considerations}
Our study does not include any personal data, as Google Search result pages are public data. Google handles circa 8.5 billion of searches per day,\footnote{https://blog.hubspot.com/marketing/google-search-statistics} so our scraping audit of 140 K daily searches provided a minimal burden. Finally, a US Federal Court has ruled that audits such as ours are legal and do not violate the Computer Fraud and Abuse Act.\footnote{\nolinkurl{https://www.aclu.org/press-releases/federal-court-rules-big-data-discrimination-studies-do-not-violate-federal-anti}}  



\begin{acks}
We are grateful to members of Wellesley Cred Lab and acknowledge funding from NSF grant ISS 1751087.
\end{acks}

\bibliographystyle{ACM-Reference-Format}
\bibliography{references}


\begin{thebibliography}{41}


\ifx \showCODEN    \undefined \def \showCODEN     #1{\unskip}     \fi
\ifx \showDOI      \undefined \def \showDOI       #1{#1}\fi
\ifx \showISBNx    \undefined \def \showISBNx     #1{\unskip}     \fi
\ifx \showISBNxiii \undefined \def \showISBNxiii  #1{\unskip}     \fi
\ifx \showISSN     \undefined \def \showISSN      #1{\unskip}     \fi
\ifx \showLCCN     \undefined \def \showLCCN      #1{\unskip}     \fi
\ifx \shownote     \undefined \def \shownote      #1{#1}          \fi
\ifx \showarticletitle \undefined \def \showarticletitle #1{#1}   \fi
\ifx \showURL      \undefined \def \showURL       {\relax}        \fi
\providecommand\bibfield[2]{#2}
\providecommand\bibinfo[2]{#2}
\providecommand\natexlab[1]{#1}
\providecommand\showeprint[2][]{arXiv:#2}

\bibitem[Abernathy(2018)]%
        {abernathy2018expanding}
\bibfield{author}{\bibinfo{person}{Penelope~Muse Abernathy}.} \bibinfo{year}{2018}\natexlab{}.
\newblock \bibinfo{booktitle}{\emph{The expanding news desert}}.
\newblock \bibinfo{publisher}{Center for Innovation and Sustainability in Local Media, School of Media and~…}.
\newblock


\bibitem[Andrea~Ballatore and Sen(2017)]%
        {BallatoreLocalness2017}
\bibfield{author}{\bibinfo{person}{Mark~Graham Andrea~Ballatore} {and} \bibinfo{person}{Shilad Sen}.} \bibinfo{year}{2017}\natexlab{}.
\newblock \showarticletitle{Digital Hegemonies: The Localness of Search Engine Results}.
\newblock \bibinfo{journal}{\emph{Annals of the American Association of Geographers}} \bibinfo{volume}{107}, \bibinfo{number}{5} (\bibinfo{year}{2017}), \bibinfo{pages}{1194--1215}.
\newblock
\urldef\tempurl%
\url{https://doi.org/10.1080/24694452.2017.1308240}
\showDOI{\tempurl}
\showeprint{https://doi.org/10.1080/24694452.2017.1308240}


\bibitem[Bandy(2021)]%
        {BandyReview2021}
\bibfield{author}{\bibinfo{person}{Jack Bandy}.} \bibinfo{year}{2021}\natexlab{}.
\newblock \showarticletitle{Problematic Machine Behavior: A Systematic Literature Review of Algorithm Audits}.
\newblock \bibinfo{journal}{\emph{Proc. ACM Hum.-Comput. Interact.}} \bibinfo{volume}{5}, \bibinfo{number}{CSCW1}, Article \bibinfo{articleno}{74} (\bibinfo{date}{apr} \bibinfo{year}{2021}), \bibinfo{numpages}{34}~pages.
\newblock
\urldef\tempurl%
\url{https://doi.org/10.1145/3449148}
\showDOI{\tempurl}


\bibitem[Bandy and Hecht(2021)]%
        {GeotargetingError2021}
\bibfield{author}{\bibinfo{person}{Jack Bandy} {and} \bibinfo{person}{Brent Hecht}.} \bibinfo{year}{2021}\natexlab{}.
\newblock \showarticletitle{Errors in Geotargeted Display Advertising: Good News for Local Journalism?}
\newblock \bibinfo{journal}{\emph{Proc. ACM Hum.-Comput. Interact.}} \bibinfo{volume}{5}, \bibinfo{number}{CSCW1}, Article \bibinfo{articleno}{92} (\bibinfo{date}{apr} \bibinfo{year}{2021}), \bibinfo{numpages}{19}~pages.
\newblock
\urldef\tempurl%
\url{https://doi.org/10.1145/3449166}
\showDOI{\tempurl}


\bibitem[Cameron(2023)]%
        {cameron2023electoral}
\bibfield{author}{\bibinfo{person}{Maxwell~A Cameron}.} \bibinfo{year}{2023}\natexlab{}.
\newblock \showarticletitle{Electoral Denialism in American Democracies}.
\newblock  (\bibinfo{year}{2023}).
\newblock


\bibitem[Courtois et~al\mbox{.}(2018)]%
        {challengeBubble2018}
\bibfield{author}{\bibinfo{person}{Cédric Courtois}, \bibinfo{person}{Laura Slechten}, {and} \bibinfo{person}{Lennert Coenen}.} \bibinfo{year}{2018}\natexlab{}.
\newblock \showarticletitle{Challenging Google Search filter bubbles in social and political information: Disconforming evidence from a digital methods case study}.
\newblock \bibinfo{journal}{\emph{Telematics and Informatics}} \bibinfo{volume}{35}, \bibinfo{number}{7} (\bibinfo{year}{2018}), \bibinfo{pages}{2006--2015}.
\newblock
\showISSN{0736-5853}
\urldef\tempurl%
\url{https://doi.org/10.1016/j.tele.2018.07.004}
\showDOI{\tempurl}


\bibitem[Diakopoulos et~al\mbox{.}(2018)]%
        {diakopoulos2018vote}
\bibfield{author}{\bibinfo{person}{Nicholas Diakopoulos}, \bibinfo{person}{Daniel Trielli}, \bibinfo{person}{Jennifer Stark}, {and} \bibinfo{person}{Sean Mussenden}.} \bibinfo{year}{2018}\natexlab{}.
\newblock \showarticletitle{I vote for—how search informs our choice of candidate}.
\newblock \bibinfo{journal}{\emph{Digital Dominance: The Power of Google, Amazon, Facebook, and Apple, M. Moore and D. Tambini (Eds.)}}  \bibinfo{volume}{22} (\bibinfo{year}{2018}).
\newblock


\bibitem[Dutton et~al\mbox{.}(2017)]%
        {SearchAndPolitics2017}
\bibfield{author}{\bibinfo{person}{William Dutton}, \bibinfo{person}{Bianca Reisdorf}, \bibinfo{person}{Elizabeth Dubois}, {and} \bibinfo{person}{Grant Blank}.} \bibinfo{year}{2017}\natexlab{}.
\newblock \showarticletitle{Search and Politics: The Uses and Impacts of Search in Britain, France, Germany, Italy, Poland, Spain, and the United States}.
\newblock \bibinfo{journal}{\emph{SSRN Electronic Journal}} (\bibinfo{date}{01} \bibinfo{year}{2017}).
\newblock
\urldef\tempurl%
\url{https://doi.org/10.2139/ssrn.2960697}
\showDOI{\tempurl}


\bibitem[Duwe and Kim(2017)]%
        {DuweRecidivism2017}
\bibfield{author}{\bibinfo{person}{Grant Duwe} {and} \bibinfo{person}{KiDeuk Kim}.} \bibinfo{year}{2017}\natexlab{}.
\newblock \showarticletitle{Out With the Old and in With the New? An Empirical Comparison of Supervised Learning Algorithms to Predict Recidivism}.
\newblock \bibinfo{journal}{\emph{Criminal Justice Policy Review}} \bibinfo{volume}{28}, \bibinfo{number}{6} (\bibinfo{year}{2017}), \bibinfo{pages}{570--600}.
\newblock
\urldef\tempurl%
\url{https://doi.org/10.1177/0887403415604899}
\showDOI{\tempurl}
\showeprint{https://doi.org/10.1177/0887403415604899}


\bibitem[Enders et~al\mbox{.}(2021)]%
        {enders20212020}
\bibfield{author}{\bibinfo{person}{Adam~M Enders}, \bibinfo{person}{Joseph~E Uscinski}, \bibinfo{person}{Casey~A Klofstad}, \bibinfo{person}{Kamal Premaratne}, \bibinfo{person}{Michelle~I Seelig}, \bibinfo{person}{Stefan Wuchty}, \bibinfo{person}{Manohar~N Murthi}, {and} \bibinfo{person}{John~R Funchion}.} \bibinfo{year}{2021}\natexlab{}.
\newblock \showarticletitle{The 2020 presidential election and beliefs about fraud: Continuity or change?}
\newblock \bibinfo{journal}{\emph{Electoral studies}}  \bibinfo{volume}{72} (\bibinfo{year}{2021}), \bibinfo{pages}{102366}.
\newblock


\bibitem[Epstein and Robertson(2015)]%
        {EpsteinSEME2015}
\bibfield{author}{\bibinfo{person}{Robert Epstein} {and} \bibinfo{person}{Ronald~E. Robertson}.} \bibinfo{year}{2015}\natexlab{}.
\newblock \showarticletitle{The search engine manipulation effect (SEME) and its possible impact on the outcomes of elections}.
\newblock \bibinfo{journal}{\emph{Proceedings of the National Academy of Sciences}} \bibinfo{volume}{112}, \bibinfo{number}{33} (\bibinfo{year}{2015}), \bibinfo{pages}{E4512--E4521}.
\newblock
\urldef\tempurl%
\url{https://doi.org/10.1073/pnas.1419828112}
\showDOI{\tempurl}
\showeprint{https://www.pnas.org/doi/pdf/10.1073/pnas.1419828112}


\bibitem[Epstein et~al\mbox{.}(2017)]%
        {EpsteinSupressingSEME2017}
\bibfield{author}{\bibinfo{person}{Robert Epstein}, \bibinfo{person}{Ronald~E. Robertson}, \bibinfo{person}{David Lazer}, {and} \bibinfo{person}{Christo Wilson}.} \bibinfo{year}{2017}\natexlab{}.
\newblock \showarticletitle{Suppressing the Search Engine Manipulation Effect (SEME)}.
\newblock \bibinfo{journal}{\emph{Proc. ACM Hum.-Comput. Interact.}} \bibinfo{volume}{1}, \bibinfo{number}{CSCW}, Article \bibinfo{articleno}{42} (\bibinfo{date}{dec} \bibinfo{year}{2017}), \bibinfo{numpages}{22}~pages.
\newblock
\urldef\tempurl%
\url{https://doi.org/10.1145/3134677}
\showDOI{\tempurl}


\bibitem[Fahey(2023)]%
        {fahey2023big}
\bibfield{author}{\bibinfo{person}{James~J Fahey}.} \bibinfo{year}{2023}\natexlab{}.
\newblock \showarticletitle{The big lie: Expressive responding and misperceptions in the United States}.
\newblock \bibinfo{journal}{\emph{Journal of Experimental Political Science}} \bibinfo{volume}{10}, \bibinfo{number}{2} (\bibinfo{year}{2023}), \bibinfo{pages}{267--278}.
\newblock


\bibitem[Fischer et~al\mbox{.}(2020)]%
        {FischerLocalNews2020}
\bibfield{author}{\bibinfo{person}{Sean Fischer}, \bibinfo{person}{Kokil Jaidka}, {and} \bibinfo{person}{Yphtach Lelkes}.} \bibinfo{year}{2020}\natexlab{}.
\newblock \showarticletitle{Auditing local news presence on Google News}.
\newblock \bibinfo{journal}{\emph{Nature human behaviour}} \bibinfo{volume}{4}, \bibinfo{number}{12} (\bibinfo{date}{December} \bibinfo{year}{2020}), \bibinfo{pages}{1236—1244}.
\newblock
\showISSN{2397-3374}
\urldef\tempurl%
\url{https://doi.org/10.1038/s41562-020-00954-0}
\showDOI{\tempurl}


\bibitem[Gillespie(2010)]%
        {PoliticsofPlatforms2010}
\bibfield{author}{\bibinfo{person}{Tarleton Gillespie}.} \bibinfo{year}{2010}\natexlab{}.
\newblock \showarticletitle{The politics of ‘platforms’}.
\newblock \bibinfo{journal}{\emph{New Media \& Society}} \bibinfo{volume}{12}, \bibinfo{number}{3} (\bibinfo{year}{2010}), \bibinfo{pages}{347--364}.
\newblock
\urldef\tempurl%
\url{https://doi.org/10.1177/1461444809342738}
\showDOI{\tempurl}
\showeprint{https://doi.org/10.1177/1461444809342738}


\bibitem[Haim et~al\mbox{.}(2018)]%
        {HaimNewsBubble2018}
\bibfield{author}{\bibinfo{person}{Mario Haim}, \bibinfo{person}{Andreas Graefe}, {and} \bibinfo{person}{Hans-Bernd Brosius}.} \bibinfo{year}{2018}\natexlab{}.
\newblock \showarticletitle{Burst of the Filter Bubble?: Effects of personalization on the diversity of Google News}.
\newblock \bibinfo{journal}{\emph{Digital Journalism}}  \bibinfo{volume}{6} (\bibinfo{date}{03} \bibinfo{year}{2018}), \bibinfo{pages}{330--343}.
\newblock
\urldef\tempurl%
\url{https://doi.org/10.1080/21670811.2017.1338145}
\showDOI{\tempurl}


\bibitem[Hannak et~al\mbox{.}(2013)]%
        {HannakPersonalization2013}
\bibfield{author}{\bibinfo{person}{Aniko Hannak}, \bibinfo{person}{Piotr Sapiezynski}, \bibinfo{person}{Arash Molavi~Kakhki}, \bibinfo{person}{Balachander Krishnamurthy}, \bibinfo{person}{David Lazer}, \bibinfo{person}{Alan Mislove}, {and} \bibinfo{person}{Christo Wilson}.} \bibinfo{year}{2013}\natexlab{}.
\newblock \showarticletitle{Measuring Personalization of Web Search}. In \bibinfo{booktitle}{\emph{Proceedings of the 22nd International Conference on World Wide Web}} (Rio de Janeiro, Brazil) \emph{(\bibinfo{series}{WWW '13})}. \bibinfo{publisher}{Association for Computing Machinery}, \bibinfo{address}{New York, NY, USA}, \bibinfo{pages}{527–538}.
\newblock
\showISBNx{9781450320351}
\urldef\tempurl%
\url{https://doi.org/10.1145/2488388.2488435}
\showDOI{\tempurl}


\bibitem[Hu et~al\mbox{.}(2019)]%
        {HuPartisonSnippets2019}
\bibfield{author}{\bibinfo{person}{Desheng Hu}, \bibinfo{person}{Shan Jiang}, \bibinfo{person}{Ronald E.~Robertson}, {and} \bibinfo{person}{Christo Wilson}.} \bibinfo{year}{2019}\natexlab{}.
\newblock \showarticletitle{Auditing the Partisanship of Google Search Snippets}. In \bibinfo{booktitle}{\emph{The World Wide Web Conference}} (San Francisco, CA, USA) \emph{(\bibinfo{series}{WWW '19})}. \bibinfo{publisher}{Association for Computing Machinery}, \bibinfo{address}{New York, NY, USA}, \bibinfo{pages}{693–704}.
\newblock
\showISBNx{9781450366748}
\urldef\tempurl%
\url{https://doi.org/10.1145/3308558.3313654}
\showDOI{\tempurl}


\bibitem[James(2000)]%
        {james2000militias}
\bibfield{author}{\bibinfo{person}{Nigel James}.} \bibinfo{year}{2000}\natexlab{}.
\newblock \showarticletitle{Militias, the Patriot movement, and the internet: the ideology of conspiracism}.
\newblock \bibinfo{journal}{\emph{The Sociological Review}} \bibinfo{volume}{48}, \bibinfo{number}{2\_suppl} (\bibinfo{year}{2000}), \bibinfo{pages}{63--92}.
\newblock


\bibitem[Johnson et~al\mbox{.}(2016)]%
        {JohnsonGeotaggedLocalness2016}
\bibfield{author}{\bibinfo{person}{Isaac~L. Johnson}, \bibinfo{person}{Subhasree Sengupta}, \bibinfo{person}{Johannes Sch\"{o}ning}, {and} \bibinfo{person}{Brent Hecht}.} \bibinfo{year}{2016}\natexlab{}.
\newblock \showarticletitle{The Geography and Importance of Localness in Geotagged Social Media}. In \bibinfo{booktitle}{\emph{Proceedings of the 2016 CHI Conference on Human Factors in Computing Systems}} (San Jose, California, USA) \emph{(\bibinfo{series}{CHI '16})}. \bibinfo{publisher}{Association for Computing Machinery}, \bibinfo{address}{New York, NY, USA}, \bibinfo{pages}{515–526}.
\newblock
\showISBNx{9781450333627}
\urldef\tempurl%
\url{https://doi.org/10.1145/2858036.2858122}
\showDOI{\tempurl}


\bibitem[Kawakami et~al\mbox{.}(2020)]%
        {kawakami2020media}
\bibfield{author}{\bibinfo{person}{Anna Kawakami}, \bibinfo{person}{Khonzodakhon Umarova}, {and} \bibinfo{person}{Eni Mustafaraj}.} \bibinfo{year}{2020}\natexlab{}.
\newblock \showarticletitle{The Media Coverage of the 2020 US Presidential Election Candidates through the Lens of Google's Top Stories}. In \bibinfo{booktitle}{\emph{Proceedings of the International AAAI Conference on Web and Social Media}}, Vol.~\bibinfo{volume}{14}. \bibinfo{pages}{868--877}.
\newblock


\bibitem[Kliman-Silver et~al\mbox{.}(2015)]%
        {LocationLocationLocation2015}
\bibfield{author}{\bibinfo{person}{Chloe Kliman-Silver}, \bibinfo{person}{Aniko Hannak}, \bibinfo{person}{David Lazer}, \bibinfo{person}{Christo Wilson}, {and} \bibinfo{person}{Alan Mislove}.} \bibinfo{year}{2015}\natexlab{}.
\newblock \showarticletitle{Location, Location, Location: The Impact of Geolocation on Web Search Personalization}. In \bibinfo{booktitle}{\emph{Proceedings of the 2015 Internet Measurement Conference}} (Tokyo, Japan) \emph{(\bibinfo{series}{IMC '15})}. \bibinfo{publisher}{Association for Computing Machinery}, \bibinfo{address}{New York, NY, USA}, \bibinfo{pages}{121–127}.
\newblock
\showISBNx{9781450338486}
\urldef\tempurl%
\url{https://doi.org/10.1145/2815675.2815714}
\showDOI{\tempurl}


\bibitem[Le et~al\mbox{.}(2019)]%
        {Le2019GoogleNews}
\bibfield{author}{\bibinfo{person}{Huyen Le}, \bibinfo{person}{Raven Maragh}, \bibinfo{person}{Brian Ekdale}, \bibinfo{person}{Andrew High}, \bibinfo{person}{Timothy Havens}, {and} \bibinfo{person}{Zubair Shafiq}.} \bibinfo{year}{2019}\natexlab{}.
\newblock \showarticletitle{Measuring Political Personalization of Google News Search}. In \bibinfo{booktitle}{\emph{The World Wide Web Conference}} (San Francisco, CA, USA) \emph{(\bibinfo{series}{WWW '19})}. \bibinfo{publisher}{Association for Computing Machinery}, \bibinfo{address}{New York, NY, USA}, \bibinfo{pages}{2957–2963}.
\newblock
\showISBNx{9781450366748}
\urldef\tempurl%
\url{https://doi.org/10.1145/3308558.3313682}
\showDOI{\tempurl}


\bibitem[Lurie and Mulligan(2021)]%
        {LurieBreakdown2021}
\bibfield{author}{\bibinfo{person}{Emma Lurie} {and} \bibinfo{person}{Deirdre~K. Mulligan}.} \bibinfo{year}{2021}\natexlab{}.
\newblock \showarticletitle{Searching for Representation: {A} sociotechnical audit of googling for members of {U.S.} Congress}.
\newblock \bibinfo{journal}{\emph{CoRR}}  \bibinfo{volume}{abs/2109.07012} (\bibinfo{year}{2021}).
\newblock
\showeprint[arXiv]{2109.07012}
\urldef\tempurl%
\url{https://arxiv.org/abs/2109.07012}
\showURL{%
\tempurl}


\bibitem[Lurie and Mustafaraj({[n.\,d.]})]%
        {LurieTopStories}
\bibfield{author}{\bibinfo{person}{Emma Lurie} {and} \bibinfo{person}{Eni Mustafaraj}.} \bibinfo{year}{[n.\,d.]}\natexlab{}.
\newblock \showarticletitle{Opening Up the Black Box: Auditing Google’s Top Stories Algorithm}.
\newblock \bibinfo{journal}{\emph{Proceedings of the ... International Florida Artificial Intelligence Research Society Conference}}  \bibinfo{volume}{32} (\bibinfo{year}{[n.\,d.]}).
\newblock
\urldef\tempurl%
\url{https://par.nsf.gov/biblio/10101277}
\showURL{%
\tempurl}


\bibitem[Metaxa et~al\mbox{.}(2019)]%
        {MetaxaSearchMedia}
\bibfield{author}{\bibinfo{person}{Dana\"{e} Metaxa}, \bibinfo{person}{Joon~Sung Park}, \bibinfo{person}{James~A. Landay}, {and} \bibinfo{person}{Jeff Hancock}.} \bibinfo{year}{2019}\natexlab{}.
\newblock \showarticletitle{Search Media and Elections: A Longitudinal Investigation of Political Search Results}.
\newblock \bibinfo{journal}{\emph{Proc. ACM Hum.-Comput. Interact.}} \bibinfo{volume}{3}, \bibinfo{number}{CSCW}, Article \bibinfo{articleno}{129} (\bibinfo{date}{nov} \bibinfo{year}{2019}), \bibinfo{numpages}{17}~pages.
\newblock
\urldef\tempurl%
\url{https://doi.org/10.1145/3359231}
\showDOI{\tempurl}


\bibitem[Metaxas and Pruksachatkun(2017)]%
        {Metaxas2017ManipulationOS}
\bibfield{author}{\bibinfo{person}{Panagiotis~Takis Metaxas} {and} \bibinfo{person}{Yada Pruksachatkun}.} \bibinfo{year}{2017}\natexlab{}.
\newblock \showarticletitle{Manipulation of Search Engine Results during the 2016 US Congressional Elections}.
\newblock
\urldef\tempurl%
\url{https://api.semanticscholar.org/CorpusID:11867830}
\showURL{%
\tempurl}


\bibitem[Mustafaraj et~al\mbox{.}(2020)]%
        {VoterCenterAudits2020}
\bibfield{author}{\bibinfo{person}{Eni Mustafaraj}, \bibinfo{person}{Emma Lurie}, {and} \bibinfo{person}{Claire Devine}.} \bibinfo{year}{2020}\natexlab{}.
\newblock \showarticletitle{The Case for Voter-Centered Audits of Search Engines during Political Elections}. In \bibinfo{booktitle}{\emph{Proceedings of the 2020 Conference on Fairness, Accountability, and Transparency}} (Barcelona, Spain) \emph{(\bibinfo{series}{FAT* '20})}. \bibinfo{publisher}{Association for Computing Machinery}, \bibinfo{address}{New York, NY, USA}, \bibinfo{pages}{559–569}.
\newblock
\showISBNx{9781450369367}
\urldef\tempurl%
\url{https://doi.org/10.1145/3351095.3372835}
\showDOI{\tempurl}


\bibitem[Pan et~al\mbox{.}(2007)]%
        {Pan2007}
\bibfield{author}{\bibinfo{person}{Bing Pan}, \bibinfo{person}{Helene Hembrooke}, \bibinfo{person}{Thorsten Joachims}, \bibinfo{person}{Lori Lorigo}, \bibinfo{person}{Geri Gay}, {and} \bibinfo{person}{Laura Granka}.} \bibinfo{year}{2007}\natexlab{}.
\newblock \showarticletitle{In Google We Trust: Users’ Decisions on Rank, Position, and Relevance}.
\newblock \bibinfo{journal}{\emph{Journal of Computer-Mediated Communication}} \bibinfo{volume}{12}, \bibinfo{number}{3} (\bibinfo{year}{2007}), \bibinfo{pages}{801--823}.
\newblock
\urldef\tempurl%
\url{https://doi.org/10.1111/j.1083-6101.2007.00351.x}
\showDOI{\tempurl}
\showeprint{https://onlinelibrary.wiley.com/doi/pdf/10.1111/j.1083-6101.2007.00351.x}


\bibitem[Puschmann(2019)]%
        {beyondTheBubbleGerman2019}
\bibfield{author}{\bibinfo{person}{Cornelius Puschmann}.} \bibinfo{year}{2019}\natexlab{}.
\newblock \showarticletitle{Beyond the Bubble: Assessing the Diversity of Political Search Results}.
\newblock \bibinfo{journal}{\emph{Digital Journalism}} \bibinfo{volume}{7}, \bibinfo{number}{6} (\bibinfo{year}{2019}), \bibinfo{pages}{824--843}.
\newblock
\urldef\tempurl%
\url{https://doi.org/10.1080/21670811.2018.1539626}
\showDOI{\tempurl}
\showeprint{https://doi.org/10.1080/21670811.2018.1539626}


\bibitem[Robertson et~al\mbox{.}(2023)]%
        {RobertsonUsers2023}
\bibfield{author}{\bibinfo{person}{Ronald Robertson}, \bibinfo{person}{Jon Green}, \bibinfo{person}{Damian Ruck}, \bibinfo{person}{Katherine Ognyanova}, \bibinfo{person}{Christo Wilson}, {and} \bibinfo{person}{David Lazer}.} \bibinfo{year}{2023}\natexlab{}.
\newblock \showarticletitle{Users choose to engage with more partisan news than they are exposed to on Google Search}.
\newblock \bibinfo{journal}{\emph{Nature}}  \bibinfo{volume}{618} (\bibinfo{date}{05} \bibinfo{year}{2023}), \bibinfo{pages}{1--7}.
\newblock
\urldef\tempurl%
\url{https://doi.org/10.1038/s41586-023-06078-5}
\showDOI{\tempurl}


\bibitem[Robertson et~al\mbox{.}(2018a)]%
        {PAB2018}
\bibfield{author}{\bibinfo{person}{Ronald~E. Robertson}, \bibinfo{person}{Shan Jiang}, \bibinfo{person}{Kenneth Joseph}, \bibinfo{person}{Lisa Friedland}, \bibinfo{person}{David Lazer}, {and} \bibinfo{person}{Christo Wilson}.} \bibinfo{year}{2018}\natexlab{a}.
\newblock \showarticletitle{Auditing Partisan Audience Bias within Google Search}.
\newblock \bibinfo{journal}{\emph{Proc. ACM Hum.-Comput. Interact.}} \bibinfo{volume}{2}, \bibinfo{number}{CSCW}, Article \bibinfo{articleno}{148} (\bibinfo{date}{nov} \bibinfo{year}{2018}), \bibinfo{numpages}{22}~pages.
\newblock
\urldef\tempurl%
\url{https://doi.org/10.1145/3274417}
\showDOI{\tempurl}


\bibitem[Robertson et~al\mbox{.}(2018b)]%
        {PersonalizationCompositionRobertson2018}
\bibfield{author}{\bibinfo{person}{Ronald~E. Robertson}, \bibinfo{person}{David Lazer}, {and} \bibinfo{person}{Christo Wilson}.} \bibinfo{year}{2018}\natexlab{b}.
\newblock \showarticletitle{Auditing the Personalization and Composition of Politically-Related Search Engine Results Pages}. In \bibinfo{booktitle}{\emph{Proceedings of the 2018 World Wide Web Conference}} (Lyon, France) \emph{(\bibinfo{series}{WWW '18})}. \bibinfo{publisher}{International World Wide Web Conferences Steering Committee}, \bibinfo{address}{Republic and Canton of Geneva, CHE}, \bibinfo{pages}{955–965}.
\newblock
\showISBNx{9781450356398}
\urldef\tempurl%
\url{https://doi.org/10.1145/3178876.3186143}
\showDOI{\tempurl}


\bibitem[Rosenblum and Muirhead(2019)]%
        {rosenblum2019lot}
\bibfield{author}{\bibinfo{person}{Nancy~L Rosenblum} {and} \bibinfo{person}{Russell Muirhead}.} \bibinfo{year}{2019}\natexlab{}.
\newblock \bibinfo{booktitle}{\emph{A lot of people are saying: The new conspiracism and the assault on democracy}}.
\newblock \bibinfo{publisher}{Princeton University Press}.
\newblock


\bibitem[Sandvig et~al\mbox{.}(2014)]%
        {Sandvig2014AuditingA}
\bibfield{author}{\bibinfo{person}{Christian Sandvig}, \bibinfo{person}{Kevin Hamilton}, \bibinfo{person}{Karrie Karahalios}, {and} \bibinfo{person}{C{\'e}dric Langbort}.} \bibinfo{year}{2014}\natexlab{}.
\newblock \showarticletitle{Auditing Algorithms : Research Methods for Detecting Discrimination on Internet Platforms}.
\newblock
\urldef\tempurl%
\url{https://api.semanticscholar.org/CorpusID:15686114}
\showURL{%
\tempurl}


\bibitem[Sen et~al\mbox{.}(2015)]%
        {SenGeoprovenance2015}
\bibfield{author}{\bibinfo{person}{Shilad~W. Sen}, \bibinfo{person}{Heather Ford}, \bibinfo{person}{David~R. Musicant}, \bibinfo{person}{Mark Graham}, \bibinfo{person}{Os Keyes}, {and} \bibinfo{person}{Brent Hecht}.} \bibinfo{year}{2015}\natexlab{}.
\newblock \showarticletitle{Barriers to the Localness of Volunteered Geographic Information}. In \bibinfo{booktitle}{\emph{Proceedings of the 33rd Annual ACM Conference on Human Factors in Computing Systems}} (Seoul, Republic of Korea) \emph{(\bibinfo{series}{CHI '15})}. \bibinfo{publisher}{Association for Computing Machinery}, \bibinfo{address}{New York, NY, USA}, \bibinfo{pages}{197–206}.
\newblock
\showISBNx{9781450331456}
\urldef\tempurl%
\url{https://doi.org/10.1145/2702123.2702170}
\showDOI{\tempurl}


\bibitem[Shelby et~al\mbox{.}(2023)]%
        {SociotechnicalHarms2023}
\bibfield{author}{\bibinfo{person}{Renee Shelby}, \bibinfo{person}{Shalaleh Rismani}, \bibinfo{person}{Kathryn Henne}, \bibinfo{person}{AJung Moon}, \bibinfo{person}{Negar Rostamzadeh}, \bibinfo{person}{Paul Nicholas}, \bibinfo{person}{N'Mah Yilla-Akbari}, \bibinfo{person}{Jess Gallegos}, \bibinfo{person}{Andrew Smart}, \bibinfo{person}{Emilio Garcia}, {and} \bibinfo{person}{Gurleen Virk}.} \bibinfo{year}{2023}\natexlab{}.
\newblock \showarticletitle{Sociotechnical Harms of Algorithmic Systems: Scoping a Taxonomy for Harm Reduction}. In \bibinfo{booktitle}{\emph{Proceedings of the 2023 AAAI/ACM Conference on AI, Ethics, and Society}} (Montreal, Canada) \emph{(\bibinfo{series}{AIES '23})}. \bibinfo{publisher}{ACM}, \bibinfo{address}{New York, NY, USA}, \bibinfo{pages}{723–741}.
\newblock
\showISBNx{9798400702310}
\urldef\tempurl%
\url{https://doi.org/10.1145/3600211.3604673}
\showDOI{\tempurl}


\bibitem[Stewart~III(2023)]%
        {stewart2023public}
\bibfield{author}{\bibinfo{person}{Charles Stewart~III}.} \bibinfo{year}{2023}\natexlab{}.
\newblock \showarticletitle{Public Opinion Roots of Election Denialism}.
\newblock \bibinfo{journal}{\emph{SSRN}} (\bibinfo{year}{2023}).
\newblock
\urldef\tempurl%
\url{http://dx.doi.org/10.2139/ssrn.4318153}
\showURL{%
\tempurl}


\bibitem[Trielli and Diakopoulos(2019)]%
        {SearchAsNewsCurator2019}
\bibfield{author}{\bibinfo{person}{Daniel Trielli} {and} \bibinfo{person}{Nicholas Diakopoulos}.} \bibinfo{year}{2019}\natexlab{}.
\newblock \showarticletitle{Search as News Curator: The Role of Google in Shaping Attention to News Information}. In \bibinfo{booktitle}{\emph{Proceedings of the 2019 CHI Conference on Human Factors in Computing Systems}} (Glasgow, Scotland Uk) \emph{(\bibinfo{series}{CHI '19})}. \bibinfo{publisher}{Association for Computing Machinery}, \bibinfo{address}{New York, NY, USA}, \bibinfo{pages}{1–15}.
\newblock
\showISBNx{9781450359702}
\urldef\tempurl%
\url{https://doi.org/10.1145/3290605.3300683}
\showDOI{\tempurl}


\bibitem[Trielli and Diakopoulos(2022)]%
        {PartisanSearchBehavior2022}
\bibfield{author}{\bibinfo{person}{Daniel Trielli} {and} \bibinfo{person}{Nicholas Diakopoulos}.} \bibinfo{year}{2022}\natexlab{}.
\newblock \showarticletitle{Partisan search behavior and Google results in the 2018 U.S. midterm elections}.
\newblock \bibinfo{journal}{\emph{Information, Communication \& Society}} \bibinfo{volume}{25}, \bibinfo{number}{1} (\bibinfo{year}{2022}), \bibinfo{pages}{145--161}.
\newblock
\urldef\tempurl%
\url{https://doi.org/10.1080/1369118X.2020.1764605}
\showDOI{\tempurl}
\showeprint{https://doi.org/10.1080/1369118X.2020.1764605}


\bibitem[Urman et~al\mbox{.}(2022)]%
        {UrmanChange2022}
\bibfield{author}{\bibinfo{person}{Aleksandra Urman}, \bibinfo{person}{Mykola Makhortykh}, {and} \bibinfo{person}{Roberto Ulloa}.} \bibinfo{year}{2022}\natexlab{}.
\newblock \showarticletitle{The Matter of Chance: Auditing Web Search Results Related to the 2020 U.S. Presidential Primary Elections Across Six Search Engines}.
\newblock \bibinfo{journal}{\emph{Social Science Computer Review}} \bibinfo{volume}{40}, \bibinfo{number}{5} (\bibinfo{year}{2022}), \bibinfo{pages}{1323--1339}.
\newblock
\urldef\tempurl%
\url{https://doi.org/10.1177/08944393211006863}
\showDOI{\tempurl}
\showeprint{https://doi.org/10.1177/08944393211006863}


\end{thebibliography}

\appendix

\end{document}